\documentclass[preprintnumbers,preprint,prd]{revtex4}
\usepackage{graphicx} 

\begin{document}

\preprint{BU-HEPP-02-01}

\title{Continuum Moment Equations on the Lattice}

\author{Walter Wilcox}
\affiliation{Physics Department, Baylor University, Waco, TX 76798-7316}

\begin{abstract}
An analysis is given as to why one can not
directly evaluate continuum moment equations, i.e., equations 
involving powers of the position
variable times charge, current, or energy/momentum operators,
on the lattice. I examine two cases: a three
point function evaluation of the nucleon magnetic moment
and a four point function (charge overlap)
evaluation of the pseudoscalar charge radius.
\end{abstract}


\maketitle

\begin{center}
\large
I. INTRODUCTION
\end{center}

Beautiful hadronic moment equations 
can be derived from continuum 
field theory three and four point matrix elements\cite{one}. 
It is very tempting to try to implement these equations 
directly in lattice simulations. A number
of papers in fact have used them, 
assumed them to be true, or examined their consequences
in lattice calculations\cite{onea,six,three,two}.
Physical quantities considered have been charge radii\cite{onea,six},
magnetic moments\cite{three,two} and quark total angular momentum
\cite{three}. However, these type of equations share one crucial 
feature in their derivation: a derivative with respect 
to momentum transfer evaluated at zero 
momentum. This last step can not be reproduced on the 
lattice because of the finite 
momentums available there, so the question arises as to the 
validity of such continuum-derived 
expressions evalauted on the lattice. We will 
examine two such expressions 
in this paper and will see that the continuum expectations 
and the lattice reality can differ markedly.

\begin{center}
\large
II. TWO EXAMPLES
\end{center}

Let us recap the situation for one such specific case, an
expression for nucleon magnetic moments
given in Refs.\cite{three,two}. Both two
and three point functions appear in this expression. 
The time ordered two point function, using the proton
interpolation field, $\chi_{\alpha}^{p}(x)$, is 
(understood $\alpha$, $\alpha'$ sums)
\begin{equation}
G_{pp}(t;\vec{p},\Gamma)  \equiv 
 \displaystyle{\sum_{\vec{x}}} 
	e^{-i\vec{p}\cdot\vec{x}}	
\Gamma_{\alpha'\alpha} 
\langle{\rm vac}|T\left(\chi_{\alpha}^{p}(x)  
	\overline{\chi}_{\alpha'}^{p}(0)\right)|{\rm vac}\rangle ,
\label{eq1}
\end{equation}
where $\Gamma_{4}\equiv \frac{1}{2}\left(\begin{array}{cc}
       I & ~~~~0 \\
       0 & ~~~~0
\end{array}\right) $ and $\Gamma_{k}\equiv \frac{1}{2}\left(\begin{array}{cc}
      \sigma_{k} & ~~~~0 \\
       0 & ~~~~0
\end{array}\right).$
The long Euclidean time limit of this expression for $\Gamma=\Gamma_{4}$ is given by
\begin{equation}
G_{pp}(t;\vec{p},\Gamma_{4}) \stackrel{t\,\gg 1}{\longrightarrow}
\frac{E_{p}+m_{N}}{2E_{p}}\frac{|Z|^{2}a^{6}}{(2\kappa)^{3}}e^{-E_{p}t},
\label{eq2}
\end{equation}
where $m_{N}$ is the (dimensionless) nucleon mass, 
\lq\lq a" is the lattice
spacing, and $\kappa = 1/(2(m+4))$, In addition,
$Z$ is the normalization factor, $({\rm
vac}|\chi_{\alpha}^{p; cont}(0)|\vec{p},s) = Zu_{\alpha}
(\vec{p},s)$, where $u_{\alpha}(\vec{p},s)$ is the free Dirac spinor,
$|\vec{p},s)$ is a continuum proton state, and $\chi_{\alpha}^{p; cont}(x)$ is the
continuum interpolation field. The three point function we need, which uses the
conserved vector current,
$J_{\mu}(x)$, is
\begin{equation}
	 G_{pJ_{\mu}p}(t_{2},t_{1};\vec{p},{\vec{p}\,'},\Gamma)  
	  \equiv 
	-i\displaystyle{ \sum_{\vec{x}_{2},\vec{x}_{1}}} 
	e^{-i \vec{p} \cdot\vec{x}_{2}}	
        e^{i\vec{q}\cdot\vec{x}_{1}} 
	 \Gamma_{\alpha'\alpha} \langle{\rm vac}|
		T\left(\chi_{\alpha}^{p}(x_{2})
	J_{\mu}(x_{1})\overline{\chi}_{\alpha'}^{p}(0)\right)|{\rm vac}\rangle,  
\label{eq3}
\end{equation}
which has the long Euclidean time limit,
\begin{equation}
G_{pJ_{j}p}(t_{2},t_{1};0,-\vec{q},\Gamma_{k})
\stackrel{(t_{2}-t_{1}),t_{1}\,\gg 1}{\longrightarrow}
\frac{1}{2E_{\ell}}\frac{|Z|^{2}a^{6}}{(2\kappa)^{3}}
e^{-m_{N}(t_{2}-t_{1})}e^{-E_{\ell}t_{1}}
\epsilon_{ijk}(\vec{q}_{\ell})_{i}G_{m}(Q^{2}_{\ell}). \label{eq4}
\end{equation}
The Minkowski four momentum transfer squared is given
by $Q^{2}_{\ell} = 2m_{N}(E_{\ell}-m_{N})$, we are
assuming continuum dispersion, $E_{\ell}^{2}=m_{N}^{2}+\vec{q}_{\ell}^{\,2}$, 
and $(\vec{q}_{\ell})_{i}=\frac{\pi \ell}{N},\,\, \ell = 0, \pm 1, \pm 2,
\dots , N$ in a given momentum direction for a square spatial lattice of size
$N_{s}=(2N)^{3}$. (All these considerations can be reformulated for a lattice 
with an odd number of spatial sites in given directions, but no fundamentally
different conclusions or observations results.) Taking a \lq\lq continuum" 
derivative of $G_{pJ_{j}p}(t_{2},t_{1};0,-\vec{q},\Gamma_{k})$ with respect to
$(\vec{q}_{\ell})_{i}$, evaluated at zero momentum, gives us another three point
function, which we will define as
\begin{equation}
G_{pJ_{j}p}(t_{2},t_{1};(\vec{x}_{1})_{i} ,\Gamma_{k}) \equiv
\displaystyle{
\sum_{\vec{x}_{2},\vec{x}_{1}}} (\Gamma_{k})_{\alpha'\alpha} 
\langle{\rm vac}|
\chi^{p}_{\alpha}(x_{2})(\vec{x}_{1})_{i}J_{j}(x_{1})
\overline{\chi}^{p}_{\alpha'}(0)|
{\rm vac}\rangle. \label{eq5}
\end{equation}
Setting this quantity equal to the continuum derivative of 
Eq.(\ref{eq4}) then results in
\begin{equation}
\left. \frac{G_{pJ_{j}p}(t_{2},t_{1};(\vec{x}_{1})_{i} ,\Gamma_{k})}
{G_{pp}(t_{2};\vec{p},\Gamma_{4})}\right|_{D}
\stackrel{(t_{2}-t_{1}),t_{1}\,\gg 1}{\longrightarrow}
\epsilon_{ijk}\frac{G_{m}(0)}{2m_{N}}. \label{eq6}
\end{equation}
The \lq\lq D" notation reminds us that this result follows from using 
a momentum derivative evaluated at zero momentum.

Ref.\cite{two} justified this procedure in the following way. One can
imagine first taking the spatial momentum derivative of the continuum analog of
$G_{pJ_{j}p}(t_{2},t_{1};0,-\vec{q},\Gamma_{k})$, evaluated at $\vec{q}=0$, and
dividing by the continuum two point function. One then transcribes this result
into lattice language by changing the continuum matrix elements into lattice ones
and making the appropriate substitutions for the spatial integration
and the various fields. It was found there that this procedure
resulted in a lattice measurement giving unrealistically small neutron
and proton magnetic moments, $G_{m}(0)$. In particular, there was
a downward trend in the data for smaller quark mass. The results in
\cite{three} are similar, although the values are larger.

In order to understand why this equation fails on the lattice, let
us expand the position variable in terms of the momentum eigenstates
of the periodic lattice. In one dimension we have the 
discrete completeness statement
\begin{equation}
\frac{1}{2N}\sum_{n=-N+1}^{N}e^{-iq_{\ell}n} = \delta_{\ell,0}. \label{eq7}
\end{equation}
On a periodic lattice with an even number
of sites, choosing an origin forces one side or the other of
the lattice to have one extra spatial site. For this purpose,
one needs to define a function which represents a linear function
everywhere except at the extra site, $n=N$, where, because
it may be considered equally distant from the origin in either direction,
we will take it to be zero. (This is the same function used 
in the numerical evaluations in \cite{three,two}.) Thus,
\begin{equation}
 F(n) = \left\{ \begin{array}{ll}
           n & \mbox{ $,n \ne N$} \\
           0 & \mbox{ $,n=N$}
\end{array}
\right.  \label{eq8}
\end{equation}
We expand this in terms of the momentum eigenfunctions, $e^{iq_{\ell}n}$,
\begin{equation}
F(n) =\sum_{l=-N+1}^{N}C_{\ell}\,e^{iq_{\ell}n}. \label{eq9}
\end{equation}
Using Eq.(\ref{eq7}), this gives 
\begin{equation}
C_{\ell} =\frac{1}{2N}\sum_{n=-N+1}^{N-1}\,F(n)\,e^{-iq_{\ell}n}. \label{eq10}
\end{equation}
Summing this finite series in the usual way by
multiplying both sides by a phase factor, $e^{-iq_{\ell}}$, and shifting the
summation limits (note that $e^{\pm iq_{\ell} N} = (-1)^{\ell}$), one finds that
\begin{equation}
C_{\ell} =  \left\{ \begin{array}{ll}
\frac{i}{2}(-1)^{\ell} \cot(q_{\ell}/2) & \mbox{$, \ell \ne 0$} \\
0 & \mbox{$,\ell=0$}
\end{array}\right. \label{eq11}
\end{equation}
This leads to
\begin{equation}
\left. \frac{G_{pJ_{j}p}(t_{2},t_{1};(\vec{x}_{1})_{i} ,\Gamma_{k})}
{G_{pp}(t_{2};\vec{p},\Gamma_{4})}\right|_{S}
\stackrel{(t_{2}-t_{1}),t_{1}\,\gg 1}{\longrightarrow}
- \frac{e^{m_{N}\,t_{1}}}{4m_{N}}\epsilon_{ijk}\sum_{\ell\ne 0}(-1)^{\ell}
\cot(q_{\ell}/2)\,
\frac{e^{-E_{\ell}\,t_{1}}}{E_{\ell}}
(\vec{q}_{\ell})_{i} G_{m}(Q^{2}_{\ell}). \label{eq12}
\end{equation}
The \lq\lq S" notation now reminds us that this result follows from 
explicitly performing the lattice sum. The coordinate $(\vec{x}_{1})_{i}$, 
when inserted in Eq.(\ref{eq5}),projects over the lattice momentums with a 
function given by Eq.(\ref{eq11}).

The leading terms in Eq.(\ref{eq12}) define
what we will term the extreme Euclidean time limit
(EETL) on the $t_{1}$ variable as,
\begin{equation}
\left. \frac{G_{pJ_{j}p}(t_{2},t_{1};(\vec{x}_{1})_{i} ,\Gamma_{k})}
{G_{pp}(t_{2};\vec{p},\Gamma_{4})}\right|_{S}
\stackrel{EETL}{\longrightarrow}
\epsilon_{ijk}
\frac{e^{(m_{N}-E_{1})\,t_{1}}}{E_{1}} G_{m}(Q^{2}_{1}), \label{eq13}
\end{equation}
where we have approximated $\sin((\vec{q}_{1})_{i}/2) \approx (\vec{q}_{1})_{i}/2$.
Eq.(\ref{eq13}) is not suitable to measure the magnetic moment on the lattice. 
At fixed finite $(\vec{q}_{1})_{i}$, the signal involves only the lowest nonzero 
component of $G_{m}(Q^{2}_{1})$, and is not time independent.

It is not a contradiction that Eqs.(\ref{eq6}) and (\ref{eq13}) 
disagree with one another even in the $q_{1}\rightarrow 0$ (or $N\rightarrow 
\infty$) limit. If one formed a discrete {\it lattice} derivative for the left side
of  Eq.(\ref{eq6}), equivalent to simply evaluating Eq.(\ref{eq4}) at the lowest 
spatial momentum and dividing by that momentum, one would obtain
a result consistent with (\ref{eq6}) in the $q_{1}\rightarrow 0$ limit.
Of course, one may always use external field methods to consistently
extract magnetic moments. Then one is effectively taking 
derivatives with respect to the external field rather than 
momentum to isolate the coupling.

\begin{figure}
\includegraphics[scale=1.0]{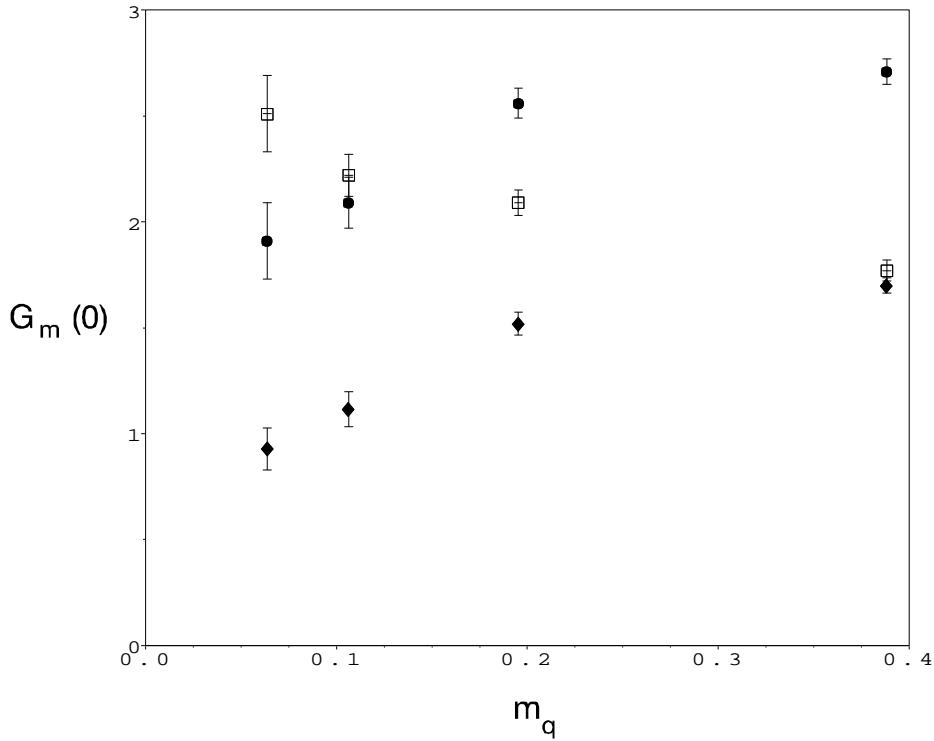}
\caption{The magnetic moment of the proton at four values of
$m_{q}$, extracted three different ways on a $16^{3}\times 24$ lattice
from the data of Ref.\cite{two}. Square symbols are extrapolated zero momentum 
form factors. The diamonds are extracted values assuming Eq.(\ref{eq6}) 
were valid. Circles show the result one would find for Eq.(\ref{eq6}) 
if the lowest momentum EETL limit in Eq.(\ref{eq13}) was dominant in the
momentum sum in Eq.(\ref{eq12}).}
\end{figure}

Fig.\ 1 shows measurements of the lattice proton magnetic
form factor at zero momentum transfer on a $16^{3}\times 24$ lattice. The open 
square symbols are extrapolated from nonzero momemtum\cite{two}. A measurement 
assuming Eq.(\ref{eq6}) yields the solid diamond symbols, which are trending downward 
as a function of decreasing quark mass. The results in Ref.\cite{three} are 
similar. Using Eq.(\ref{eq13}) one can test the extent to which the lowest momentum is
contributing to the result from (\ref{eq12}) because one has separate data for
$G_{m}(Q_{1}^{2})$ at the lowest momentum transfer\footnote{$G_{m}(Q_{1}^{2})$ is 
obtained by averaging over spatial momentums.}. The solid circles
show the result one would find for Eq.(\ref{eq6}) if the lowest momentum EETL 
limit in Eq.(\ref{eq13}) was dominant in the momentum sum. The trend is also 
downward for smaller quark mass, which is a result of the decreasing value of
$G_{m}(Q_{1}^{2})$ as well as the greater exponential suppression from the
$e^{(m_{N}-E)t}$ factor. ($t=7\frac{1}{2}$ for the data from Ref.\cite{two}). The
qualitative behaviors are remarkably similar for these two Eq.(\ref{eq6}) 
measurements, although significant cancellation is probably occurring for the 
diamond data from the $(-1)^{\ell}$ factor in Eq.(\ref{eq12}). It was speculated
in \cite{two,three} that the downward trend possibly reflected the fact that the
nucleon was not well contained in the lattice volume. The Fig.\ 1 data strongly
suggests that the downward trend in the diamond symbols is instead the result of
two factors, decreasing $G_{m}(Q_{\ell}^{2})$ and greater $e^{(m_{N}-E)t}$
suppression, similar to the EETL case\footnote{That the extrapolated proton 
$G_{m}(0)$ value is increasing while the measured $G_{m}(Q_{\ell}^{2})$ 
values are decreasing for smaller $m_{q}$ is not a contradiction, but perfectly 
consistent with the assumed dipole form of the fits in Ref.\cite{two}, which 
has two independent parameters, $G_{m}(0)$ and the dipole mass, $m_{D}$.}.

As another example of a continuum moment equation on the lattice, consider the
charge overlap measurement of the pion charge radius. Time separated measurements
of charge overlap matrix elements were first considered in lattice calculations in 
Ref.\cite{four}.

We will start with the results derived in Ref\cite{five}. Using
$u,d$ flavor conserved lattice charge densities, $\rho^{u,d}(x)$, one has
for $t_{3}\gg t_{1,2}\gg1$,
\begin{equation}
\frac{\sum_{\vec{x}_{3},\vec{z}}<0|\phi^{\dagger}(x_{3})T\left(\rho^{u}(x_{2})
\rho^{d}(x_{1})\right)\phi(z)|0>}{\sum_{\vec{x}_{3},\vec{z}}<0|
\phi^{\dagger}(x_{3})\phi(z)|0>} \longrightarrow
{\cal P}^{ud}_{\pi}(\vec{r},t),
\label{eq14}
\end{equation}
where $z =(\vec{z},0)$ and $\phi$ is a charged pion interpolation field. ${\cal
P}^{ud}_{\pi}(\vec{r},t)$ can be written using space and time translational 
invariance between the zero momentum pion states ($\vec{r}
\equiv
\vec{x}_{2}-\vec{x}_{1}, t\equiv t_{2}-t_{1}$) as
\begin{equation}
{\cal P}^{ud}_{\pi}(\vec{r},t) = <\pi(\vec{0})|T(\rho^{u}(r)
\rho^{d}(0))|\pi(\vec{0})>.
\label{eq15}
\end{equation}
The discrete Fourier transform is defined to be,
\begin{eqnarray}
{\cal Q}(\vec{q}^{\,2},t)\equiv N_{s}\sum_{\vec{r}}e^{-i\vec{q}\cdot
\vec{r}}{\cal P}^{ud}_{\pi}(\vec{r},t).
\label{eq16}
\end{eqnarray}
It is important to let $t\gg 1$ in Euclidean space in order to 
damp out the contributions of higher mass intermediate states when a complete set
of states is inserted between the charge densities
in (\ref{eq15}). One then has
\begin{eqnarray}
{\cal Q}(\vec{q}^{\,2},t)
\stackrel{t\gg 1}{\longrightarrow}
\frac{(E_{q}+m_{\pi})^{2}}{4E_{q}m_{\pi}} F_{\pi}^{2}(Q^{2})e^{(m_{\pi}-E_{q})t},
\label{eq17}
\end{eqnarray}
where $Q^{2}=2m_{\pi}(E_{q}-m_{\pi})$ and $F_{\pi}(Q^{2})$ is 
the pion form factor.

A continuum derivative of Eq.(\ref{eq16}) with respect to $\vec{q}^{\,2}$ 
at zero momentum forms the quantity 
\begin{equation}
{\cal R}^{2}_{\pi}(t)\equiv N_{s}\sum_{\vec{r}}\vec{r}^{\,2}
{\cal P}^{ud}_{\pi}(\vec{r},t).
\label{eq18}
\end{equation}
Using Eqs.(\ref{eq16}),(\ref{eq17}) and (\ref{eq18}) and following a 
procedure similar to the above for the magnetic moment, we obtain the 
continuum derivative result,
\begin{equation}
\left. {\cal R}^{2}_{\pi}(t)\right|_{D} \stackrel{t\,\gg 1}{\longrightarrow}
 2R^{2}_{u,d} +\frac{3t}{m_{\pi}},
\label{eq19}
\end{equation}
where $R^{2}_{u,d}$ is the charged pion $u,d$ quark charge radius. This is the same result
as in Ref.\cite{six} when the bag sources there are  replaced with zero momentum 
pion sources. One could imagine evaluating this expression on the lattice to try to extract the
charge radius from the time constant term, but we will see this hope is ill-founded.

In order to explain why one can not measure the charge radius from an expression
like Eq.(\ref{eq18}), let us now expand the square of the position variable, 
$\vec{r}^{\,2}$, in terms of the momentum eigenstates. For a one dimensional 
lattice we will
consider,
\begin{equation}
n^{2} =\sum_{\ell=-N+1}^{N}K_{\ell}\,e^{iq_{\ell}n}.
\label{eq20}
\end{equation}
We then find that the coefficients, $K_{\ell}$, are given by
\begin{equation}
K_{\ell} = \left\{ \begin{array}{ll}
\frac{1}{2}(-1)^{\ell}\csc^{2}(q_{\ell}/2) & \mbox{ $,\ell \ne 0$} \\
\frac{1}{3}(N^{2}+\frac{1}{2})  & \mbox{ $,\ell = 0$}
\end{array}\right.
\label{eq21}
\end{equation}
In a three dimensional context, the quantities on the right in (\ref{eq21}) are
multiplied by zero momentum Kronecker deltas in the transverse directions.

Using Eqs.(\ref{eq20}) and (\ref{eq21}) in (\ref{eq18}) and inserting a complete
set, one finds that
\begin{equation}
\left. {\cal R}^{2}_{\pi}(t)\right|_{S} \stackrel{t\,\gg 1}{\longrightarrow}
(N^{2}+\frac{1}{2})+\frac{3}{2}\sum_{\ell \ne 0}(-1)^{\ell}\csc^{2}
(q_{\ell}/2)
\frac{(E_{\ell}+m_{\pi})^{2}}{4m_{\pi}E_{\ell}} F_{\pi}^{2}(Q^{2}_{\ell})
e^{(m_{\pi}-E_{\ell})t}.
\label{eq22}
\end{equation}
We can define a discrete {\it lattice} charge radius as
\begin{equation}
(R^{2}_{u,d})_{L}\equiv 3\frac{(1-F^{2}_{\pi}(Q^{2}_{1}))}{q_{1}^{2}}.
\label{eq23}
\end{equation}
In the EETL, we then find,
\begin{equation}
\left. {\cal R}^{2}_{\pi}(t)\right|_{S}
\stackrel{EETL}{\longrightarrow}
(N^{2}+\frac{1}{2})-12(\frac{N^{2}}{\pi^{2}}-\frac{1}{3}(R^{2}_{u,d})_{L})
e^{(m_{\pi}-E_{1})t}.
\label{eq24}
\end{equation}
So we see that in contrast to Eq.(\ref{eq19}), the first (time independent) term is 
essentially meaningless. It is conceivable that one could extract a measurement of 
$(R^{2}_{u,d})_{L}$ from the second, time dependent term, but it would be much 
simpler to project out the form factor at the lowest lattice momentum 
in the usual manner and use Eq.(\ref{eq23}). Again, the fact that
(\ref{eq19}) and (\ref{eq24}) do not agree even for very fine lattices 
($N\rightarrow\infty$) is not a contradiction because one has not taken 
a discrete lattice derivative, but a continuum one in producing (\ref{eq19}).

\begin{center}
\large
III. SUMMARY
\end{center}

We have seen why it is not possible to
directly evaluate continuum moment equations on a periodic lattice.
Continuum moments of lattice operators in a periodic system
do not project onto good momentum and so do not isolate low momentum 
properties. The present author pointed this fact out some years ago
in the context of charged pion polarizability calculations\cite{five}. 
In a more general sense, the lesson we have learned is that in order to 
deduce continuum properties from the lattice, it is important to treat 
the lattice as a self-consistent physical system. Position functions 
are meaningful only if expanded in terms of the available momentum 
eigenstates of the system. It can be very misleading in general to try 
to take continuum field theory equations and simply \lq\lq latticize" 
them in deducing physical properties. 
  
\begin{center}
\large
IV. ACKNOWLEDGMENTS
\end{center}

This work is supported in part by NSF Grant No.\ 0070836. The Baylor University
Sabbatical Program is also gratefully acknowledged.

\end{document}